\documentstyle[12pt]{article}
\newcommand{\mq}{m_{\tilde{q}}}
\begin{document}
\vspace*{-1in}
\renewcommand{\thefootnote}{\fnsymbol{footnote}}
\begin{flushright}
CERN-TH. 7530/94 \\
hep-ph/9503264 \\
\end{flushright}
\vskip 65pt
\begin{center}
{\Large \bf \boldmath New LEP constraints on some supersymmetric
Yukawa interactions that violate $R$-parity} \\
\vspace{8mm}
{\bf Gautam Bhattacharyya\footnote{gautam@cernvm.cern.ch},
John Ellis\footnote{johne@cernvm.cern.ch}}
and {\bf K. Sridhar\footnote{sridhar@vxcern.cern.ch}}\\
\vspace{10pt}
{\bf Theory Division, CERN, \\ CH-1211, Gen\`eve 23, Switzerland.}

\vspace{80pt}
{\bf ABSTRACT}
\end{center}
We consider one-loop corrections to partial widths of the $Z$
induced by supersymmetric Yukawa interactions that violate
$R$-parity. The precise experimental values of the leptonic $Z$
partial widths bound these Yukawa couplings, with the most
interesting constraints being those on couplings involving the
$\tau$, since previous constraints on them were very mild.
\vspace{98pt}
\noindent
\begin{flushleft}
CERN-TH. 7530/94\\
February 1995\\
\end{flushleft}

\vskip 10pt
\begin{center}
Submitted to {\it Mod. Phys. Lett.} {\bf A}.
\end{center}

\setcounter{footnote}{0}
\renewcommand{\thefootnote}{\arabic{footnote}}
\vfill
\clearpage
\setcounter{page}{1}
\pagestyle{plain}

Of the many possible extensions of the Standard Model (SM),
supersymmetry is considered to be one of the most promising
candidates and, consequently, a significant amount of theoretical
and experimental effort has been devoted to looking for signals of
supersymmetry at present and future colliders. In particular, the
minimal supersymmetric extension of the Standard Model (MSSM)
\cite{mssm} has been the subject of numerous investigations.  In
addition to the usual particles of the Standard Model, the MSSM
contains their superpartners and two Higgs doublets. The gauge
structure of the MSSM essentially replicates that of the Standard
Model; there is no arbitrariness in the structure of the gauge
interactions.  This is, however, not true for the Yukawa sector of
the MSSM.  In addition to the usual Yukawa couplings of the
fermions to the Higgs (responsible for the fermion masses), other
interactions involving squarks or sleptons are possible.

\vskip 10pt

The relevant part of the superpotential containing the Yukawa
interactions involving squarks or sleptons in the MSSM is given in
terms of the chiral superfields by
\begin{equation}
    {\cal W}_{\not R} = \lambda_{ijk} L_i L_j E^c_k +
                        \lambda'_{ijk} L_i Q_j D^c_k +
                        \lambda''_{ijk} U^c_i D^c_j D^c_k
      \label{e1}
\end{equation}
where the $L_i$ and $Q_i $ are $SU(2)$-doublet lepton and quark
fields and the $E^c_i, U^c_i, D^c_i$ are singlet superfields.  In
general, there are 45 such Yukawa couplings:~ 9 each of the
$\lambda$- and $\lambda''$- types (because of the antisymmetry in
the first (last) two generation indices of the former (latter)
couplings) and 27 of the $\lambda'$-type. The couplings $\lambda$
and $\lambda'$ violate lepton ($L$) number, whereas the $\lambda''$
coupling violates baryon ($B$) number. The $B$- and $L$-violating
couplings cannot be present simultaneously, because that would lead
to very rapid proton decay.  It is possible to forbid the existence
of all three interactions in eq.(\ref{e1}) by imposing a discrete
symmetry -- $R$-parity. This discrete symmetry may be represented
by $R=(-1)^{(3B+L+2S)}$, where $S$ is the spin of the particle, so
that the usual particles of the SM have $R=1$, while their
superpartners have $R=-1$.

\vskip 10pt

The requirement that the MSSM Lagrangian be invariant under
$R$-parity is sufficient to exclude each of the interactions in
Eq.(\ref{e1}).  There is, however, no compelling theoretical
argument in favour of such a symmetry \cite{rparv}. $R$-parity
conservation is too strong a requirement to ensure proton
stability~-- the latter can simply be ensured by assuming that
either the $L$-violating or the $B$-violating couplings in
Eq.(\ref{e1}) are present, but not both.  Relaxing the requirement
of $R$-parity conservation has important implications for
supersymmetric particle searches at colliders~: a superparticle can
decay into standard particles {\it via} the $R$-parity violating
couplings in eq.(\ref{e1}).

\vskip 10pt

In this letter we examine the constraints on some of the
$L$-violating $\lambda'$-type couplings imposed by precision LEP
observables, particularly the $Z \rightarrow l^+ l^-$ partial decay
widths. There are new triangle diagrams with $Z, l^+$ and $l^-$
external lines involving $\lambda'_{ijk}$ vertices with $i =$
lepton, $j =$ quark, $k =$ squark indices or $i =$ lepton, $j =$
squark, $k =$ quark indices. Since the magnitude of the new
contribution depends on the mass of the fermion in the loop, only
$\lambda'_{i3k}$-type couplings leading to internal top quark lines
can be constrained significantly by our considerations
\footnote{For a review of previous phenomenological constraints on
$\lambda'_{ijk}$, see \cite{dimo,bgh}, and for a recent
complementary study, see \cite{bc}.}.
The constraints
we derive require these couplings to be smaller than the $SU(2)$
gauge coupling or the top quark -- Higgs Yukawa coupling if the
squarks are $\sim$ 100 GeV.

\vskip 10pt

We should also recall that there exist important cosmological
constraints \cite{cosm} on $R$-parity-violating scenarios.
Requiring that GUT-scale baryogenesis does not get washed out
imposes $\lambda''<< 10^{-7}$ generically, though these bounds are
model dependent and can be evaded \cite{dr}.  Assuming $\lambda''=
0$, the $\lambda'$ couplings cannot wash out the initial baryon
asymmetry by themselves.  However, they can do so in association
with a $B$-violating but $(B-L)$ conserving interaction, such as
sphaleron-induced non-perturbative transitions. Since these
processes conserve ${1\over3}B - L_i$ for each lepton generation,
the conservation of any one lepton generation number is sufficient
\cite{nel} to retain the initial baryon asymetry. Therefore, the
assumption that the smallest $\lambda'$-type coupling is less than
$\sim 10^{-7}$ is enough to avoid any cosmological bound on the
remaining $\lambda'$-type couplings.  Hence, we proceed with a
$B$-conserving but $L$-violating scenario, assuming $\lambda'' =
0$, some theoretical motivations for which can be found in ref.
\cite{hall}.

\vskip 10pt

Since we are only interested in the interactions generated by the
$\lambda'$ couplings, we write this part of the Lagrangian out in
terms of the component fields. In four-component Dirac notation,
the complete set of Lagrangian terms with $\lambda'$ couplings is
given by
\begin{eqnarray}
    {\cal L} =  \lambda'_{ijk} \lbrack
                \tilde{\nu}_L^i \bar{d}_R^k d_L^j +
                \tilde{d}_L^j \bar{d}_R^k \nu_L^i +
                (\tilde{d}_R^k)^* (\bar{\nu}_L^i)^c d_L^j
 \nonumber \\
                - \tilde{e}_L^i \bar{d}_R^k u_L^j -
                \tilde{u}_L^j \bar{d}_R^k e_L^i -
                (\tilde{d}_R^k)^* (\bar{e}_L^i)^c u_L^j
                                \rbrack  + \mbox{\rm h.c.}
      \label{e2}
\end{eqnarray}
In our work, we concentrate on computing the one-loop corrections
arising due to the interactions in Eq.(\ref{e2}) to the partial
widths $Z \rightarrow l^+l^-$ (where $l^-$ and $l^+$ denote a
charged lepton and the corresponding antilepton, respectively),
because these widths are very precisely determined experimentally.
Since the experimental determination of the leptonic width requires
knowledge of the hadronic width of the $Z$, it is important also to
know the effect of the $\lambda'_{i3k}$ couplings on the hadronic
channels, for example, $Z \rightarrow \bar{b}b$
\footnote{In fact, the dominant correction to the total hadronic
width comes from $Z \rightarrow \bar{b}b (\bar{d}d, \bar{s}s)$
depending upon whether the $R$-violating Yukawa coupling is
$\lambda'_{i33} (\lambda'_{i31}, \lambda'_{i32})$. We consider, of
course, only one of them at a time. Moreover, since the tree-level
SM predictions for $\Gamma_{b(d,s)}$ are same, the bounds on
$\lambda'_{i3k}$ are by and large the same irrespective of the
choice of $k$ for a common sparticle mass.} In fact, it is clear
that the effect on $\Gamma_b$ of the $i=$ lepton, $j=$ quark
(squark), $k=$ squark (quark) coupling is negligibly small, since
this vertex involves leptons in the loop, and hence gives
corrections which are, at best, a factor $m_{\tau}^2 / m_t^2$
smaller than the corrections to the leptonic width. But a sizeable
correction to $\Gamma_b$ can arise due to the $i=$ slepton, $j=$
quark, $k=$ quark vertex, since the corresponding loops involve the
top quark.

\vskip 10pt

The tree level $Z$ couplings to the left- and right-handed fermions
are given by $a_L^f$ and $a_R^f$, respectively, which appear in
\begin{equation}
M_\mu^{\rm tree} = {e \over{s_Wc_W}} \bar{f}(p^\prime) \gamma_\mu
(a_L^f P_L + a_R^f P_R) f(p).
\end{equation}
where
\begin{eqnarray}
a_L^f & = & t_3^f - Q_f s_W^2, \nonumber  \\
a_R^f & = & - Q_f s_W^2.
\end{eqnarray}
The $Z$ couplings to the charge-conjugated fermions ($f^c$)
are, therefore,
\begin{equation}
a_L^{f^c} = - a_R^f, ~~~~~~~~~~~~~~~~~ a_R^{f^c} = - a_L^f.
\end{equation}
The triangle and self-energy diagrams that we need to compute for
the correction to the processes $Z \rightarrow l^+l^-$ and $Z
\rightarrow \bar b b$ are shown in Figs.~1a and 1b, respectively.
We compute these diagrams in terms of the Passarino-Veltman $B$-
and $C$-functions
\cite{pasvel}, corresponding to the two- and three-point integrals.
For the purposes of presentation, we give explicit expressions for
one diagram of Fig.~1a, {\it viz.}, the top-induced loop diagram
contributing to $Z \rightarrow l^+ l^-$. It is easy to see that
only the coupling of the left-handed leptons to the $Z$ is modified
by this diagram, so that the amplitude due to the new contribution
is
\begin{equation}
\label{e5}
M_{\mu}^{(i)} = {N_c\over 16 \pi^2} {e {{\lambda'}^2} \over s_Wc_W}
\bar l (p^{\prime}) \gamma_{\mu} A_i l(p) ,
\end{equation}
where $i = 1,2,3$ and $N_c$ is the colour factor.
The $A_i$'s are given by,
\begin{eqnarray}
\label{e6}
A_1 &=&  \left[a_L^{t^c} m_t^2 C_0 -
a_R^{t^c} \lbrace m_Z^2 (C_{22}-C_{23}) + (d-2)C_{24} \rbrace
\right] P_L , \nonumber \\
A_2 &=&-2  c_{\tilde d} {\tilde C}_{24} P_L, \nonumber \\
A_3 &=&  a_L^l B_1 P_L.
\end{eqnarray}
Here $A_{1,2}$ denote the contributions from the first and the
second triangle diagrams, and the contributions of the two
self-energy diagrams are jointly denoted by $A_3$.  In $A_2$, we
use $\tilde{C}_{24}$ to distinguish it from the $C_{24}$ appearing
in $A_1$, as the structures of the propagators for the two triangle
diagrams are different. In the expression for $A_2$, $c_{\tilde d}$
refers to the coupling of the $\tilde{d}_R^k$-squark to the $Z$.
We point out that the contributions from the individual diagrams
are divergent, namely $C_{24}$ in $A_1$, ${\tilde C}_{24}$ in $A_2$
and $B_1$ in $A_3$. But the divergence cancels when these
amplitudes are added, and we are left with a finite correction.
There is also a finite correction from the other set of triangle
and self-energy diagrams in Fig.~1a, calculated analogously to
eq.(\ref{e6}) with appropriate modifications. Together these finite
parts make a new contribution to the partial width $Z \rightarrow
l^+l^-$, which we denote by $\delta \Gamma_l$.  We have evaluated
the $B$- and $C$-functions required using the code developed in
ref.\cite{amit}, cross-checking the results by using the standard
Feynman parametrisation of the two- and three-point functions and
then integrating them numerically.

\vskip 10pt

To provide intuition into our numerical results, we also present
analytic expressions valid in the limit $m_t, \mq \gg m_Z$, for an
arbitrary value of the ratio $x \equiv m_t^2/\mq^2$.  The sum of
the $A_i$'s in eq.(\ref{e6}) is given in this limit by
\begin{equation}
\sum_{j=1}^3 A_j
= \biggl\lbrack (a_L^{t^c}-a_R^{t^c}) \eta_2(x) +
{{m_Z^2}\over{3 m_t^2}} \left\{a_R^{t^c} \eta_1(x) +
c_{\tilde{d}} \eta_3(x)\right\}
\biggr\rbrack,
\label{e6sim}
\end{equation}
where
$\eta_1, \eta_2$ and $\eta_3$ are given by
\begin{eqnarray}
\eta_1(x) & = & {{-11x+18x^2-9x^3+2x^4}\over{6(1-x)^4}}
- {{x\ln x}\over{(1-x)^4}} \simeq 0~ (x \rightarrow 0),  \nonumber \\
\eta_2(x) & = & -{{x}\over{1-x}} - {{x\ln x}\over{(1-x)^2}}
\simeq 0~ (x \rightarrow 0) , \\
\eta_3(x) & = & {{2x-9x^2+18x^3-11x^4}\over{6(1-x)^4}}
+{{x^4\ln x}\over{(1-x)^4}} \simeq 0~ (x \rightarrow 0).\nonumber
\end{eqnarray}
Eqs (8) and (9) exhibit the expected decoupling in the limit of large
$\mq$, and match qualitatively the exact numerical results presented
later even for smaller values of $\mq$.

\vskip 10pt

We neglect left-right squark mixing, which is expected to be
significant only for the $\tilde{t}$ \cite{elru}. As noted earlier,
the diagram that contributes dominantly to $\delta\Gamma_l$ is that
with the top quark in the loop, and this diagram contains a
$b$-squark, which justifies our neglect of left-right squark
mixing.  Similarly, from a computation of the diagrams in Fig.~1b,
we obtain a modification to $\Gamma_b$, which we denote by $\delta
\Gamma_b$ \footnote {It may be noted that, for $\lambda'_{i33}$
couplings, another set of triangle and self energy diagrams with
external $b_L$- and internal $b_R$-lines in addition to the
diagrams shown in Fig. 1b has to be taken into account. As has been
pointed out, $Z \rightarrow
\bar{b}b$ also receives contributions from internal lepton and
squark lines. However, these contributions are small and are
neglected.}.  We observe that the quantity $\delta \Gamma_l$ is a
function of the squark mass, whereas $\delta \Gamma_b$ depends {\it
mainly} on the slepton mass.  Although the squark and slepton
masses are completely free parameters phenomenologically, the usual
assumption of unification at some Grand Unified scale would
normally predict the physical slepton masses to be smaller than the
squark masses: the low-energy splitting can be calculated exactly
by renormalization group evolutions from some initial conditions in
specific models. For our purposes, it is a safe approximation to
work with the slepton mass equal to half of the squark mass: our
results are insensitive to plausible variations from this
approximation.

\vskip 10pt

We compute numerically the corrections to the ratio $R_l$, defined
as:
\begin{equation}
R_l = {\Gamma_h \over \Gamma_l}  \hskip15pt (l=e,\mu,\tau).
\end{equation}
Defining the quantities $\Delta_f \ (f=l,b)$ by
\begin{equation}
\Delta_f = {\Gamma_f - \Gamma_f^{\rm SM} \over
 \Gamma_f^{\rm SM} } ,
\end{equation}
one has the following expression for the change in $R_l$:
\begin{equation}
\delta R_l \equiv R_l -R_l^{\rm SM}
\approx R_l^{\rm SM}R_b^{\rm SM}\Delta_b-
R_l^{\rm SM}\Delta_l .
\end{equation}
Using this equation and, since $\delta R_l$ always turns out
negative, comparing the 2-$\sigma$ experimental upper bound on $R_l$
with the corresponding SM prediction, we obtain the maximum allowed
value of $\lambda'_{i3k}$ as a function of the squark mass.
We use the following experimental values \cite{lep}:
$R_e = 20.850 \pm 0.067,
R_\mu = 20.824 \pm 0.059,
R_\tau = 20.749 \pm 0.070$.
The SM predictions for $R_l$ and $R_b$ are 20.786 and 0.2158
respectively for the following choice of input parameters: $m_t
=175$~GeV, $m_H =300$~GeV and $\alpha_s =0.126$ \cite{lep}.
The ratio $R_l$ is quite insensitive to these values, so our
results do not change much by varying these parameters.  Fig.~2
shows the values of $\lambda'_{\mbox{\rm max}}$ obtained as
functions of the squark mass for $e$, $\mu$ and $\tau$ final
states.  We observe that for $\mq = 100$ GeV, $\lambda'_{33k} \leq
0.45, \lambda'_{23k} \leq 0.56$ and $\lambda'_{13k} \leq 0.63$ at
the two-standard deviation level.  We observe that the previous low
energy bound on the coupling $\lambda'_{133}$ is much stronger
\footnote {We wish to point out that the upper limit of $\nu_e$
Majorana mass can be used to put constraints on
$\lambda'_{1jk}$ {\it only when} $j=k$, and not in the general case
for any $j$ and $k$ as suggested in ref.\cite{grt}.}
than the one obtained from our analysis, while the rest of our bounds
are new, with the exception of those on $\lambda'_{131}$ and
$\lambda'_{231}$. Our bounds
could be significantly improved as more data accumulate on the
$Z$-peak.  A compilation of our and previous phenomenological
bounds on $\lambda'_{i3k}$ are displayed in Table 1, where, for the
sake of consistent comparison with the previous bounds, we also
present the constraints we obtain at the one-standard deviation
level.

\begin{table}[htbp]
\begin{center}
\bigskip
\begin{tabular}{|c|c|c|c|c|c|c|c|c|c|}
\hline
\multicolumn{4}{|c|}{Previous analyses} &
\multicolumn{6}{|c|}{Present analysis} \\
\hline
$\lambda'_{13k}$ & Limits & $\lambda'_{23k}$ & Limits &
$\lambda'_{13k}$ & Limits &
$\lambda'_{23k}$ & Limits & $\lambda'_{33k}$ & Limits \\
\hline
131 & 0.26  & 231 & $0.22^*$ & 131 & 0.63 & 231 & 0.56 & 331 & 0.45 \\
    &       &     &          &   & (0.51) &   & (0.44) &   & (0.26) \\
132 &       & 232 &          & 132 & 0.63 & 232 & 0.56 & 332 & 0.45 \\
    &       &     &          &   & (0.51) &   & (0.44) &   & (0.26) \\
133 & 0.001 & 233 &          & 133 & 0.63 & 233 & 0.56 & 333 & 0.45 \\
    &       &     &          &   & (0.51) &   & (0.44) &   & (0.26) \\
\hline
\end{tabular}
\caption[] {Limits on the $\lambda'_{i3k}$ couplings from the
previous and present analyses for $\mq = 100$ GeV.  The number
marked by ($*$) in the column of previous analyses corresponds to a
2-$\sigma$ limit, while the other numbers correspond to 1-$\sigma$
limits (see \cite{bgh,grt}).  In the columns of the present
analysis, we give numbers corresponding to 2-$\sigma$ limits
followed by 1-$\sigma$ limits in brackets.}
\end{center}
\end{table}

\vskip 10pt

In passing, we note that the above interactions may also affect the
SM predictions of the forward-backward charge asymmetries
($A_{FB}^l$), the $\tau$-polarisation asymmetry ($A^\tau$) or the
left-right asymmetry ($A_{LR}$), however, in view of our
constraints, the effects are below the experimental sensitivity, as
can be inferred from the similarity of the $\lambda'$-type
$R$-parity-violating interactions to leptoquark-induced
contributions to the same processes \cite{bes}.

\vskip 10pt

Our bounds have implications for sparticle decays in models with
$R$-violation. In the past \cite{past}, sparticle decays into
leptons, particularly $\tau$ and/or $\nu_\tau$, have been discussed
from a phenomenological point of view. Our bounds imply that decays
{\it via} $\lambda'$-type couplings are unlikely to dominate over
$R$-parity conserving decays of heavier sparticles into gauginos,
at least if sparticles weigh $\sim 100$ GeV. Searches for
experimental signatures of $R$-parity violation should therefore
focus on decays of the lightest supersymmetric particle, which
could still be rapid enough to occur inside a detector.

\vskip 10pt

In conclusion: we have considered the effect of the
$R$-parity-violating couplings $\lambda'$ on the $Z$ partial
widths.  By studying the effect of the interactions induced by
these couplings on the ratio of the hadronic and leptonic widths,
we have been able to place reasonably stringent constraints on the
couplings $\lambda'$ as functions of the masses of the squarks and
sleptons.  The available data allow us to get interesting bounds
which are complementary to those previously existing, particularly
for interactions involving the $\tau$ lepton.  Our bounds could be
improved significantly with more precise data.

\vskip 10pt
\noindent{\bf Acknowledgements}
\par
We thank D.~Choudhury for useful discussions.

\newpage

\tolerance=100000
\input axodraw.sty

\vskip 4cm

\begin{picture}(240,100)
\Photon(20,50)(80,50){5}{4}
\ArrowLine(120,10)(80,50)
\ArrowLine(80,50)(120,90)
\DashLine(120,10)(120,90){5}
\ArrowLine(160,10)(120,10)
\ArrowLine(120,90)(160,90)
\LongArrow(65,39)(35,39)
\put(25,62){$Z(p-p^{\prime})$}
\put(165,90){$l(p^{\prime})$}
\put(165,7){$l(p)$}
\put(100,82){$q$}
\put(100,35){$q$}
\put(125,50){$\tilde{q}$}

\Photon(220,50)(280,50){5}{4}
\DashLine(320,10)(280,50){5}
\DashLine(280,50)(320,90){5}
\ArrowLine(320,10)(320,90)
\ArrowLine(360,10)(320,10)
\ArrowLine(320,90)(360,90)
\LongArrow(265,39)(235,39)
\put(225,62){$Z(p-p^{\prime})$}
\put(365,90){$l(p^{\prime})$}
\put(365,7){$l(p)$}
\put(300,82){$\tilde{q}$}
\put(300,35){$\tilde{q}$}
\put(325,50){$q$}
\end{picture}
\vskip 0.5cm

\vskip 5cm

\begin{picture}(240,100)
\Photon(20,50)(80,50){5}{4}
\ArrowLine(120,10)(80,50)
\ArrowLine(80,50)(120,90)
\DashCArc(100,70)(15,45,225){5}
\LongArrow(65,39)(35,39)
\put(25,62){$Z(p-p^{\prime})$}
\put(125,90){$l(p^{\prime})$}
\put(125,7){$l(p)$}
\put(105,65){$q$}
\put(83,83){$\tilde{q}$}

\Photon(220,50)(280,50){5}{4}
\ArrowLine(320,10)(280,50)
\ArrowLine(280,50)(320,90)
\DashCArc(300,30)(15,135,315){5}
\LongArrow(265,39)(235,39)
\put(225,62){$Z(p-p^{\prime})$}
\put(325,90){$l(p^{\prime})$}
\put(325,7){$l(p)$}
\put(302,35){$q$}
\put(283,3){$\tilde{q}$}
\end{picture}
\vskip 2.5cm
\centerline{\Large $q=t^c ,b$;\ $\tilde{q}=\tilde{b}, \tilde{t}^*$ }

\vfill
\begin{quote}
Figure 1a:~
The one-loop Feynman diagrams contributing to the $Z \rightarrow
l^+l^-$ vertex correction due to the $R$-parity-violating coupling
$\lambda'$.
\end{quote}

\newpage

\begin{picture}(240,100)
\Photon(20,50)(80,50){5}{4}
\ArrowLine(120,10)(80,50)
\ArrowLine(80,50)(120,90)
\DashLine(120,10)(120,90){5}
\ArrowLine(160,10)(120,10)
\ArrowLine(120,90)(160,90)
\LongArrow(65,39)(35,39)
\put(25,62){$Z(p-p^{\prime})$}
\put(165,90){$b(p^{\prime})$}
\put(165,7){$b(p)$}
\put(100,82){$q$}
\put(100,35){$q$}
\put(125,50){$\tilde{L}$}

\Photon(220,50)(280,50){5}{4}
\DashLine(320,10)(280,50){5}
\DashLine(280,50)(320,90){5}
\ArrowLine(320,10)(320,90)
\ArrowLine(360,10)(320,10)
\ArrowLine(320,90)(360,90)
\LongArrow(265,39)(235,39)
\put(225,62){$Z(p-p^{\prime})$}
\put(365,90){$b(p^{\prime})$}
\put(365,7){$b(p)$}
\put(300,82){$\tilde{L}$}
\put(300,35){$\tilde{L}$}
\put(325,50){$q$}
\end{picture}
\vskip 0.5cm

\vskip 5cm

\begin{picture}(240,100)
\Photon(20,50)(80,50){5}{4}
\ArrowLine(120,10)(80,50)
\ArrowLine(80,50)(120,90)
\DashCArc(100,70)(15,45,225){5}
\LongArrow(65,39)(35,39)
\put(25,62){$Z(p-p^{\prime})$}
\put(125,90){$b(p^{\prime})$}
\put(125,7){$b(p)$}
\put(105,65){$q$}
\put(83,83){$\tilde{L}$}

\Photon(220,50)(280,50){5}{4}
\ArrowLine(320,10)(280,50)
\ArrowLine(280,50)(320,90)
\DashCArc(300,30)(15,135,315){5}
\LongArrow(265,39)(235,39)
\put(225,62){$Z(p-p^{\prime})$}
\put(325,90){$b(p^{\prime})$}
\put(325,7){$b(p)$}
\put(302,35){$q$}
\put(283,3){$\tilde{L}$}
\end{picture}
\vskip 2.5cm
\centerline{\Large $q=t,b$;\ $\tilde{L}=\tilde{l}, \tilde{\nu_l}$ }

\vfill
\begin{quote}
Figure 1b:~
The one-loop Feynman diagrams contributing to the $Z \rightarrow
\bar b b$ vertex correction due to the $R$-parity-violating coupling
$\lambda'$.
\end{quote}

\newpage
\setcounter{figure}{1}
\begin{figure}[htb]
\vskip 8in\relax\noindent\hskip -1in\relax{\includegraphics{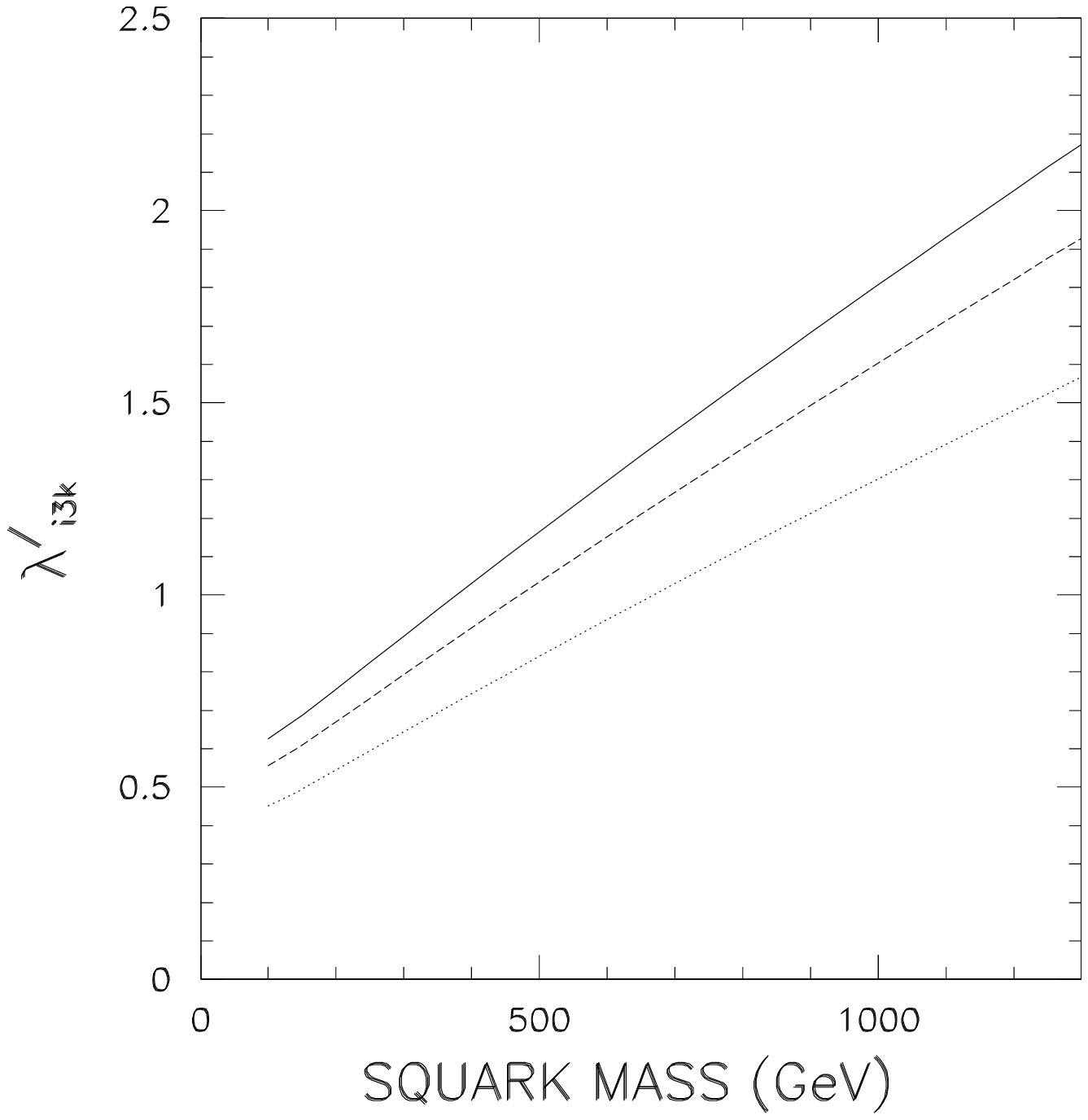}}

\vspace{-20ex}
\caption{The maximum values of the $\lambda'_{i3k}$ (2-$\sigma$
limits) as functions of the squark mass allowed by the $e, \mu$ and
$\tau$ partial widths of the $Z$ (solid, dashed and dotted lines,
corresponding to $i = 1, 2$ and 3, respectively).}
\end{figure}
\end{document}